\documentclass[prb,showpacs,floats,superscriptaddress,twocolumn,amsmath,amssymb]{revtex4-1}

\usepackage[dvipdfmx]{graphicx}
\usepackage{bm}
\usepackage{times}

\begin{document}

\preprint{Preprint}

\title{Emergent loop-nodal s$_\pm$-wave superconductivity in CeCu$_2$Si$_2$: similarities to the iron-based superconductors}%

\author{Hiroaki Ikeda}
\email{ikedah@fc.ritsumei.ac.jp}
\affiliation{Department of Physics, Ritsumeikan University, Kusatsu 525-8577, Japan}
\affiliation{Department of Physics, Kyoto University, Kyoto 606-8502, Japan}
\author{Michi-To Suzuki}
\affiliation{RIKEN Center for Emergent Matter Science (CEMS), Wako, Saitama 351-0198, Japan}
\affiliation{CCSE, Japan Atomic Energy Agency, 5-1-5 Kashiwanoha, Kashiwa, Chiba 277-8587, Japan}
\author{Ryotaro Arita}
\affiliation{RIKEN Center for Emergent Matter Science (CEMS), Wako, Saitama 351-0198, Japan}
\affiliation{Department of Applied Physics, University of Tokyo, Tokyo 113-8656, Japan}

\date{\today}%

\begin{abstract}
Heavy-fermion superconductors are prime candidates for novel electron-pairing states due to the spin-orbital coupled degrees of freedom and electron correlations. Superconductivity in CeCu$_2$Si$_2$ discovered in 1979, which is a prototype of unconventional (non-BCS) superconductors in strongly correlated electron systems, still remains unsolved. Here we provide the first report of superconductivity based on the advanced first-principles theoretical approach. We find that the promising candidate is an $s_\pm$-wave state with loop-shaped nodes on the Fermi surface, different from the widely expected line-nodal $d$-wave state. The dominant pairing glue is magnetic but high-rank octupolar fluctuations. This system shares the importance of multi-orbital degrees of freedom with the iron-based superconductors. Our findings reveal not only the long-standing puzzle in this material, but also urge us to reconsider the pairing states and mechanisms in all heavy-fermion superconductors.
\end{abstract}


\maketitle

\vfill

Some sort of intermetallic compounds, containing elements with 4f or 5f electrons, are called heavy fermion materials because the low-energy excitations can be described by heavy electrons with the large effective mass up to 1000 times the free-electron mass. In these f-electron systems, the strong electron correlation and many degrees of freedom bring about a variety of curious and interesting physical phenomena, such as complex magnetic/multipole ordering, quantum critical phenomena, and unconventional superconductivity and so on. 

The heavy-fermion material CeCu$_2$Si$_2$ is the first unconventional superconductor discovered in 1979  by F. Steglich et al. [\onlinecite{rf:Steglich}]. The specific heat coefficient $C/T\sim 0.75$J/mol$\cdot$Ce\,K$^2$ at $T_c\sim 0.5$K indicates that the heavy-fermion state has been formed by the strong electron correlation between f electrons. The BCS-like specific heat jump at $T_c$ is a clear evidence of the Cooper pairing of the correlated electrons, in other words, the gap opening in heavy-electron bands with high density of states (DOS). The relatively high $T_c$ to the effective Fermi energy $T_F\sim 10$K made a great impact on the research field since it has been considered that the strong Coulomb repulsion disturbs the superconductivity. Also, much small $T_F$, as compared with the Debye temperature $\sim 200$K, is contrary to the case in the conventional BCS theory. Since the early stage, these facts have implied that the superconductivity is not conventional. Indeed after a short time, unconventional behavior in superconducting phase was observed in some experimental works; the $T$-linear behavior at low $T$ in $C/T$ [\onlinecite{rf:Arndt,rf:Bredl}], and no coherence peak just below $T_c$ and the $T^3$ behavior in NMR relaxation rate $1/T_1$ [\onlinecite{rf:Kitaoka,rf:Ishida,rf:Fujiwara}]. These observations are inconsistent with the exponential decay in the fully-gapped $s$-wave superconductivity in the BCS theory, rather indicate line nodes on the Fermi surface as in the $d$-wave superconductivity in the subsequently discovered high-$T_c$ cuprates. Thus it has been widely expected that the line-nodal $d$-wave state is the promising candidate of the superconductivity in CeCu$_2$Si$_2$. Actually this material and the high-$T_c$ cuprates share some characteristic features~\cite{rf:Holmes,rf:Jaccard,rf:Monthoux}; non-Fermi liquid behavior above the optimal $T_c$, and the similar phase diagram in which superconductivity appears in close proximity to the antiferromagnetic (AFM) phase.

These facts imply also that the Coulomb repulsion, especially a magnetic fluctuation, is crucially important to make mobile electrons bound strongly. In the conventional BCS superconductors, the major glue for electron pairings is quanta of lattice vibration, i.e., phonon. The obtained $s$-wave order parameter has a large weight on the on-site pairing amplitude in real space. Therefore, the on-site Coulomb repulsion disturbs the on-site pairing and suppresses $T_c$. On the other hand, a plausible pairing glue in the high-$T_c$ superconductors is the AFM fluctuations driven by the Coulomb repulsion, which leads to anisotropic pairs like $d$-wave state. This anisotropic pairing state without on-site amplitude is not directly suppressed by the on-site Coulomb repulsion. This is a reason why the Coulomb repulsion can lead to the high-$T_c$ superconductivity. The superconducting pairing function is closely related to the pairing mechanism. Clarifying the pairing state is crucially important to understand the pairing mechanism. Such microscopic studies will give us some useful hints for how to raise the transition temperatures. Indeed a recent strategy for finding new high-$T_c$ materials is to increase the energy scale of dynamical spin fluctuations. It is based on the fact that the $T_c$ of Ce115, Pu115 and the cuprate systems is linearly scaled to the characteristic energy of AFM fluctuations~\cite{rf:Nakai,rf:Pines}.

On the other hand, finding new electron-pairing mechanism can offer another route to raise $T_c$, such as AFM fluctuations instead of phonon. In this sense, the finding of the iron-based superconductors~\cite{rf:Kamihara} may give us a chance to search for another mechanism, since the orbital degrees of freedom are important ingredients in these materials. In this regard, heavy-fermion superconductors can provide a playground to search for much variety of pairing mechanisms and pairing states. However, the complicated electronic band structure has prevented our understanding of nature of superconductivity so far. Under the circumstances, recent developments on the first-principles calculations based on the density-functional theory (DFT) have promoted a breakthrough on studying the electronic state in the strongly correlated electron systems. For instance, DFT+DMFT (dynamical mean-field theory) methods allow us to discuss the correlated electrons, compared to angle-resolved photoemmision spectroscopy (ARPES) results in real materials~\cite{rf:Shim}. On the other hands, as a complementary method, based on an effective model obtained from the DFT calculations, low-energy magnetic/multipole fluctuations have been investigated in the context of the hidden-order transition in URu$_2$Si$_2$ [\onlinecite{rf:Ikeda}]. 
The same approach can be a powerful tool to dissect the complex electronic state in heavy-fermion materials. 
We here provide the first report of the microscopic analysis about the superconducting gap function in CeCu$_2$Si$_2$. We find that the situation in CeCu$_2$Si$_2$ is similar to that in the iron-based superconductors, and the promising pairing state is $s_\pm$-wave state, which is in stark contrast to the widely expected $d$-wave state. The dominant pairing glue is magnetic but high-rank octupolar fluctuations. In addition we discuss that the second dome of superconductivity under high pressures~\cite{rf:Yuan} can possess the similar $s$-wave pairing, but mediated by different mechanism. 

\begin{figure}[h]
\caption{(color online). (a, b) Fermi surface colored by the Fermi velocity, (c, d) in-plane magnetic (dipole) RPA susceptibilities for $q=(q_x,q_y,0.5)$, and (e, f) a complete set of multipole susceptibilities along the high-symmetry line~\cite{rf:note}. Left figures correspond to the LDA+$U$ case, and right to the ordinary LDA case. In (a), the presence of heavy-electron sheet around $X$ point is similar to the case of the renormalized band method~\cite{rf:Zwicknagl}. Incommensurate peak positions $Q=(0.21,0.21,0.5)$ in (c) are consistent with the characteristic $Q$ vector observed in neutron scattering measurements~\cite{rf:Stockert}, while in the LDA case, (d), it is hard to enhance magnetic fluctuations even for larger interactions. In (e) we can see that octupole (Rank 3) fluctuations are remarkably enhanced in the LDA+$U$ case.  In the LDA case, (f), on the other hand, non-magnetic quadrupole/hexadecapole fluctuations are relatively enhanced. Here, susceptibilities without a peak at $X$ point are uniformly colored red.}
\includegraphics[width=0.5\textwidth]{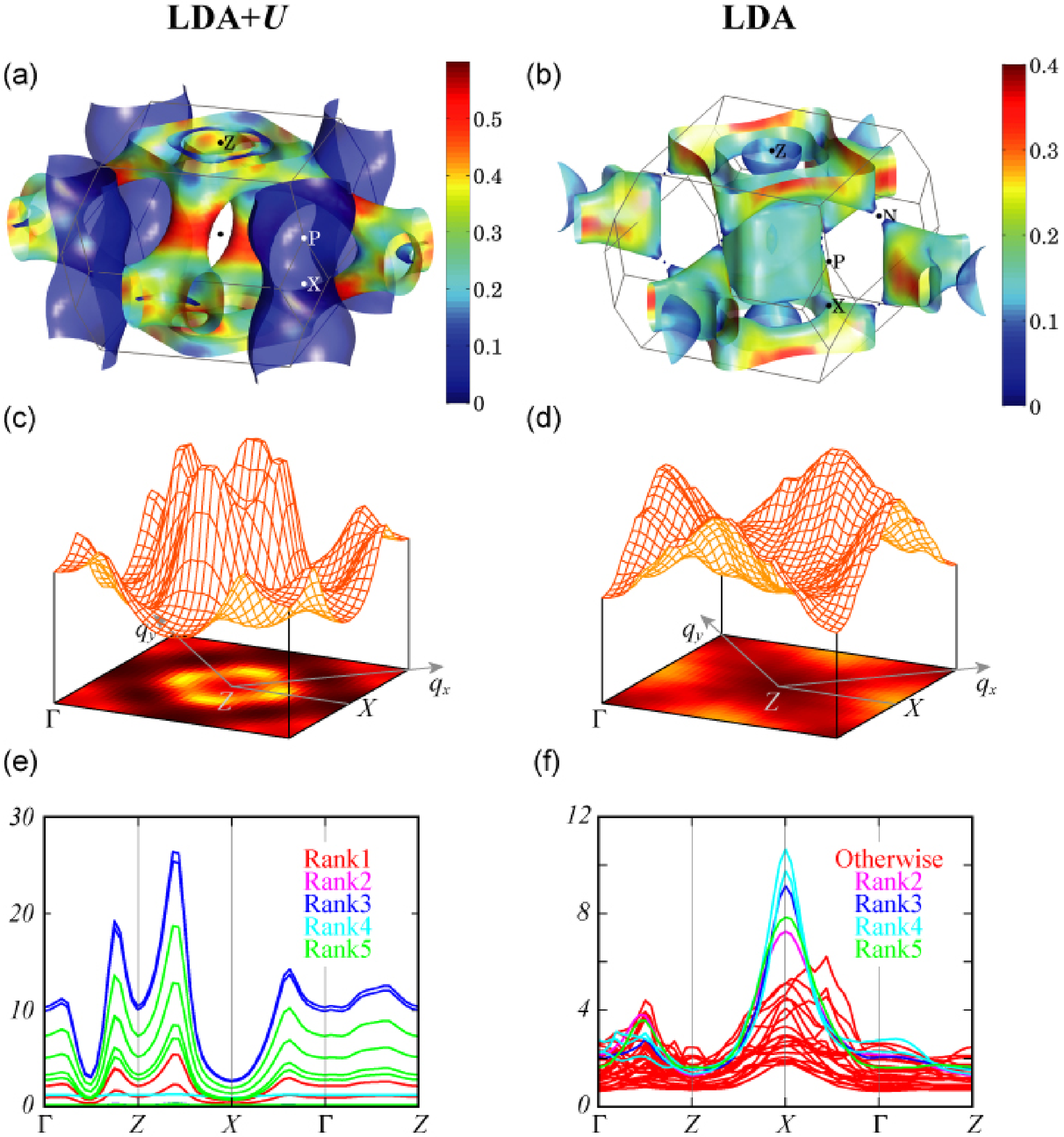}
\end{figure}

{\it The Fermi surface and magnetic/multipole fluctuations ---}
The Fermi surface topology, which is crucially important for unconventional superconductivity, can be experimentally clarified by the quantum oscillation and ARPES measurements and so on. However, the Fermi surface in CeCu$_2$Si$_2$ is not so clear yet. Theoretically, from the electronic band structure calculations, two possible Fermi surface topology have been proposed, which are clearly different each other. One is obtained by the ordinary LDA calculations~\cite{rf:Harima}, and the other by the renormalized band method~\cite{rf:Zwicknagl}. The Fermi surface in the latter is similar to Fig.1(a), which was obtained by our LDA+$U$ calculations~\cite{rf:Kittaka,rf:note01}. In the LDA Fermi surface shown by Fig.1(b), there are a cubic-like electron sheet and a tiny electron sheet around $\Gamma$ and a complex hole sheet.  In the LDA+$U$ case, there appears a corrugated-cylinder electron sheet around $X$ point with heavy effective mass and a complex hole sheet and a tiny hole ring. The band crossing the Fermi level is mainly composed of an f-orbital manifold with the total angular momentum $j=5/2$. In the LDA+$U$ calculation~\cite{rf:SM}, it is especially dominated by its azimuthal component $j_z=\pm 3/2$ mixed with a small weight of $j_z=\mp 5/2$, although the latter components may become larger by including the electron correlations~\cite{rf:Willers,rf:Pourovskii}.

First of all, let us investigate the Fermi surface topology and magnetic fluctuations. We compute low-energy magnetic fluctuations in both LDA and LDA+$U$ Fermi surfaces within the random-phase approximation (RPA).~\cite{rf:note}. In the LDA+$U$ Fermi surface shown by Fig.1(c), we find a peak structure at around incommensurate $Q=(0.21,0.21,0.5)$ due to the nesting property in the corrugated heavy-electron sheet around $X$ point, which is successfully consistent with a characteristic $Q$ vector observed in the neutron scattering measurements~\cite{rf:Stockert}. Similar trend has been discussed in a simple model~\cite{rf:Eremin}. On the contrary, a hump structure around $X$ in Fig.1(d) in the LDA case is not so much enhanced even for larger interactions. These facts imply that the Fermi surface in CeCu$_2$Si$_2$ should possess a feature of corrugated-cylinder electron sheet as observed in the LDA+$U$ and the renormalized band case.
Hence, we consider hereafter that the Fermi surface at the ambient pressure is like the LDA+$U$ Fermi surface hereafter.

Next let us study a variety of inter-orbital fluctuations~\cite{rf:note}. Generally speaking, in multi-orbital systems, there exist the orbital degrees of freedom in addition to spin degrees of freedom. The presence of strong spin-orbit coupling in the f-electron systems entangles these degrees of freedom, which leads to the so-called multipole degrees of freedom. In the $j=5/2$ multiplet with six $j_z$ components ($j_z=\pm 5/2, \pm 3/2, \pm 1/2$), there exists $6\times 6=36$ multipole degrees of freedom, which are classified into monopole (rank 0), dipole (rank 1), quadrupole (rank 2), octupole (rank 3), hexadecapole (rank 4), and dotriacontapole (rank 5) by symmetries in the group theory. Conventional charge and magnetic degrees of freedom correspond to monopole and magnetic dipole, respectively. High-rank multipoles represent a spin-orbital coupled degrees of freedom. ``Rank'' means changeable difference of total angular momentum. For example, an operator $f^\dagger_{+5/2}f_{-5/2}$, which means a change between $j_z=+5/2$ and $-5/2$, is classified into a kind of the highest rank 5. This state has been discussed as a promising hidden-order parameter in URu$_2$Si$_2$ [\onlinecite{rf:Ikeda}]. 

Fig.1(e) depicts a complete set of multipole susceptibilities in the LDA+$U$ case of CeCu$_2$Si$_2$. We can see that octupole fluctuations are dominantly enhanced, and next is rank 5 fluctuations. This is consistent with the fact that the major component of the ground-state f multiplets is $j_z=\pm 3/2$ mixed with small components of $j_z=\mp 5/2$. These high-rank fluctuations are much larger than the magnetic dipole fluctuations. This implies that the incommensurate AFM order observed in the A-type materials possesses a sizable weight of the octupole moment with the same irreducible representation. On the other hand, in the LDA case shown in Fig.1(d), it is hard to enhance magnetic fluctuations even for larger interactions. Rather, non-magnetic quadrupole/hexadecapole fluctuations become relatively large (Fig.1(f))~\cite{rf:Hattori}. Recently, L.V. Pourovskii et al. proposed a possibility of orbital transition at around the second superconducting dome under high pressures~\cite{rf:Pourovskii}. In our case this corresponds to a kind of Lifshitz transition from the LDA+$U$ Fermi surface with anisotropic f-electron charge distribution to the LDA Fermi surface with almost isotropic distribution. This implies that with applying pressure, incommensurate AFM fluctuations are suppressed, but non-magnetic orbital fluctuations are enhanced instead. It is an interesting open problem how these non-magnetic fluctuations are related to valence fluctuations~\cite{rf:Holmes,rf:Rueff,rf:Kobayashi,rf:Miyake,rf:Yamaoka}, which is one of hotly-debated issues in heavy-electron systems. 

\begin{figure*}[!t]
\caption{(color online). (a) Schematic pressure-temperature phase diagram in CeCu$_2$Si$_2$ [\onlinecite{rf:Yuan}]. (b-e) Superconducting gap structures obtained in Eq.(1). The bottom figures are the top view or the side view. (b) $d_{x^2-y^2}$-wave ($\sim\!\!\cos(2k_x)-\cos(2k_y)$) and (c) $s_\pm$-wave ($\sim\!\!\cos(2k_x)+\cos(2k_y)$) obtained in the LDA+$U$ Fermi surface. (d) Another $s_\pm$-wave state in the LDA case, $\sim\!\!\cos(k_x)\cos(k_y)\cos(k_z)$, and (e) $d_{xy}$-wave ($\sim\!\!\sin(k_x)\sin(k_y)\cos(k_z)$). Two $s_\pm$-wave states, (c) and (d), are the promising superconducting state in CeCu$_2$Si$_2$, different from the widely believed $d$-wave state. The dominant pairing interactions come from the octupole fluctuations in (c) and the quadrupole/hexadecapole fluctuations in (d). In addition, these two $s_\pm$-wave state with different pairing mechanisms may correspond to two distinct superconducting phases observed under pressures in (a).}
\includegraphics[width=0.8\textwidth]{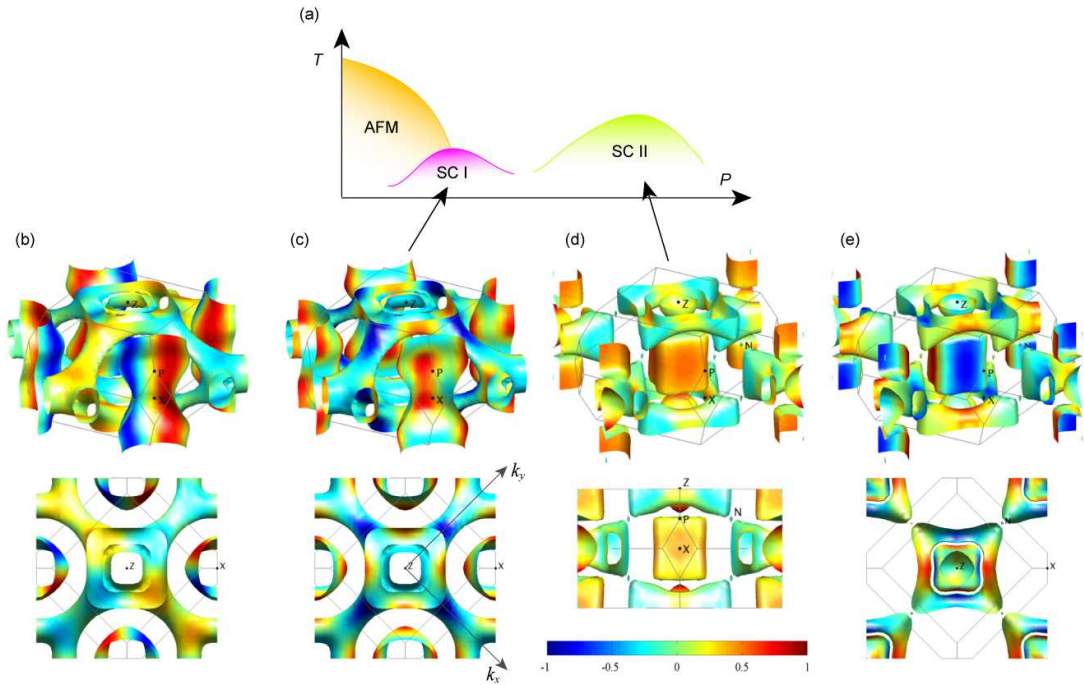}
\end{figure*}

{\it Superconducting pairing symmetry ---}
Here, let us discuss what kind of superconductivity can emerge by multipole fluctuations obtained above. As usual, we evaluate the linearized gap equation in the multiorbital systems~\cite{rf:Ikeda2}, 
\begin{align}
\begin{split}
\Delta_{\ell m}(k)=-\lambda \sum_{k'}\sum_{\ell'\ell''m'm''} V_{\ell\ell',m'm}(k-k')~~~~~~~~~~ \\
{\cal G}_{\ell'\ell''}(k'){\cal G}_{m'm''}(k')^* \Delta_{\ell''m''}(k'),
\end{split}
\end{align}
where $\Delta_{\ell m}(k)$ represents the pair function, and $V_{\ell m,\ell'm'}(q)$ is the pairing interaction,
\begin{align}
\begin{split}
&V_{\ell m,\ell'm'}(\boldsymbol{q}) = \Gamma_{\ell m,\ell'm'}^0+ \\
&~~~~~~~~~~\sum_{\ell_1m_1\ell_2m_2}\Gamma_{\ell m,\ell_1m_1}^0\chi_{\ell_1 m_1,\ell_2 m_2}(\boldsymbol{q})\Gamma_{\ell_2 m_2,\ell'm'}^0.
\end{split}
\end{align}
Information of multipole fluctuations in this system is contained in $\chi_{\ell m,\ell' m'}(\boldsymbol{q})$, which is $\chi^0_{\ell m,\ell' m'}(\boldsymbol{q})$ for the second-order perturbation and $\chi^{\rm RPA}_{\ell m,\ell' m'}(\boldsymbol{q})$ for the RPA. 
First of all, in the RPA susceptibilities, we obtain a $d_{x^2-y^2}$ wave state ($\lambda=1.73$) and an $s_\pm$ wave ($\lambda=1.43$) for the LDA+$U$ Fermi surface, which are due to the remarkable octupole fluctuations (Fig.1(e)), while an $s_\pm$ wave ($\lambda=1.16$) and a $d_{xy}$ wave ($\lambda=0.56$) for the LDA Fermi surface due to the comparably enhanced non-magnetic fluctuations (Fig.1(f)). 
The leading $d_{x^2-y^2}$ wave state obtained in the LDA+$U$ case is the widely believed line-nodal $d$-wave state. However, the sub-leading $s_\pm$ wave also has a large eigenvalue, and it is the leading state in the LDA case. In general, within the RPA, magnetic susceptibilities are relatively too much enhanced as compared with non-magnetic (charge/orbital) ones. Therefore, the second-order perturbation can provide a complementary information. Indeed it will be more appropriate in this material, since this material has only weak correlation between magnetic fluctuations and transition temperatures. 

In the second-order perturbation, we obtain the $s_\pm$ wave symmetry as the leading pairing state, independent of the Fermi-surface topology. The leading $s_\pm$ wave has $\lambda=1.06$ and the sub-leading $d_{x^2-y^2}$ wave has $\lambda=0.97$ for the LDA+$U$ case, while the leading pairing state remains the $s_\pm$ wave ($\lambda=0.75$) for the LDA case. As shown in Fig.2(c), the $s_\pm$-wave state obtained in the LDA+$U$ case has a complicated structure with loop-shaped nodes on the Fermi surface. The similar loop-nodal $s_\pm$ wave state has been discussed in the iron-based superconductor BaFe$_2$(As$_{1-x}$P$_x$)$_2$ [\onlinecite{rf:Yamashita}]. Although this loop-nodal $s_\pm$-wave state has still nodal excitations on the corrugated heavy-electron sheet, it should be noted that a gap size on a flat part is much smaller than that on a convex part. With a slight mixture of the on-site pairing due to intrinsic attractive on-site interaction by higher-order vertex corrections~\cite{rf:Onari}, the loop-nodes can be lifted since this nodal feature is not symmetry-protected. In this case, the corrugated heavy-electron sheet becomes fully-gaped, and only the light-hole sheet possesses loop-nodes. This is also the case in the LDA case. Thus, the advanced first-principles calculations are indicative that the $s_\pm$-wave pairing with the loop nodes on the light Fermi surface is the promising superconducting state in CeCu$_2$Si$_2$.

Experimentally, recent specific-heat measurements~\cite{rf:Kittaka} show the exponential behavior below 60mK and the $H$-linear dependence under the magnetic field $H$. It has been indicated that these features can be explained by multi-gapped $s$-wave state along with the $T$-linear behavior above 60mK. No coherence peak in the NMR $1/T_1$ does not contradict an $s_\pm$-wave state. Since the peak structure observed in the neutron measurements~\cite{rf:Stockert2} is located at around $\sim 2\Delta$ with the estimated gap $\Delta$, it is not so clear whether this is the resonance peak. Rather, such weak peak structure seems to support our high-rank multipole-fluctuation mechanism. Thus the $s_\pm$-wave pairing state can be the most probable candidate in this material. Then, why such possibility has been missed? First, before the discovery of iron-based superconductors, it has been considered that power-law behavior in a wide temperature range is a strong evidence of nodal structure, rather than multi-gapped $s$-wave. Second, it has not been so seriously taken due to strong sample dependence that its extrapolation to 0K of a power-law fit in the specific heat is negative. Thus, recent improvement of sample quality and low-temperature techniques has triggered reconsideration of superconductivity in this material.

Our present results strongly depend on the Fermi-surface topology. Therefore, it is crucial to clarify the Fermi surface experimentally. In addition, whether the loop nodes exist on the hole sheet with light effective mass requires further studies at low temperatures in terms of several experiments sensitive to low-energy excitations, such as magnetic penetration depth and thermal conductivity and so on. 

Finally, we realize a possibility that curious two superconducting domes (Fig.2(a)) observed under high pressures~\cite{rf:Yuan} can be explained by two types of $s_\pm$-wave states along with a kind of Lifshitz transition from LDA+$U$ Fermi surface to the LDA Fermi surface. These pairing states are mediated by two different fluctuations; one is magnetic octupolar fluctuations and another is non-magnetic hexadecapolar fluctuations. This implies that the multi-orbital degrees of freedom are crucially important. This feature is very similar to two superconducting domes recently discussed in the iron-based superconductors~\cite{rf:Mukuda}. Thus, our study is also helpful to the iron-based superconductors. Our findings of the $s_\pm$-wave state in CeCu$_2$Si$_2$ will stimulate us to reconsider the pairing states and mechanisms in all of heavy-fermion superconductors in detail.

\begin{acknowledgements}
We acknowledge S. Kittaka, T. Sakakibara, C. Geibel, and F. Steglich for a recent data of the specific heat, and thank J. Schmalian, P. Thalmeier, Y. Kitaoka, K. Ishida, K. Machida, Y. Matsuda, and T. Shibauchi for helpful discussion. This work was partially supported by Grants-in-Aid for Scientific Research (KAKENHI) from Japan Society for the Promotion of Science (JSPS).
\end{acknowledgements}



\begin{thebibliography}{99}
\bibitem{rf:note01} Note that our LDA+$U$ approach is not in a spin-polarized but a paramagnetic case. The former case is often useful to study electronic structure in the localized f-electron systems, while the latter involves important f-band shift in itinerant f-electron systems. This method possesses a tendency that f-electron charge distribution becomes anisotropic, as compared with the LDA calculations~\cite{rf:Suzuki}. In our case, $j_z=\pm 3/2$ component is dominant. This can be compatible with a clear crystal-field splitting observed in resistivity measurements at the ambient pressure.
\bibitem{rf:note00} 
In this letter, we set on-site interactions $U=U'=1.3$eV and $J=J'=0$ expect for the RPA calculations in the LDA+$U$ case, where smaller $U=U'=0.24$eV was used to avoid a divergence of the RPA susceptibilities. Since inclusion of $J=J'$ does not change momentum dependence of susceptibilities so much, we set $J=J'=0$ for simplicity. It should be noted that these $U$ values are smaller than those used in the LDA+$U$ calculations themselves. This discrepancy can be resolved by considering the Kondo physics, such as DMFT calculations, because interactions in the present RPA calculations should be considered as the quasi-particle interactions, which are renormalized and smaller than the bare values. Calculations of multipole susceptibilities based on the LDA+DMFT are one of interesting issues in the future. However, it can be expected that momentum dependence of susceptibilities, which is important for unconventional superconductivity, is not so strongly affected by the Kondo physics.
\end{thebibliography}
\end{document}